\documentclass[twoside]{article}
\usepackage{amssymb}
\usepackage{qic,epsfig}

\def\IC{{\mathbb C}}
\def\IR{{\mathbb R}}

\def\cH{{\mathcal H}}

\def\cH{{\mathcal H}}

\def\cK{{\mathcal K}}

\def\cA{{\mathcal A}}

\def\tr{{\rm tr}}
\def\({\left(}
\def\){\right)}
\def\diag{{\rm diag}\,}

\def\int{{\bf Int}}

\def\ra{{\rangle}}
\def\la{{\langle}}
\def\qed{$\square$\,}

\begin{document}
%\openup 1 \jot

\runninghead{Recovery in quantum error correction for general noise without measurement}
            {C-K Li, M. Nakahara, Y-T Poon, N-S Sze, H. Tomita}

\normalsize\textlineskip
\thispagestyle{empty}
\setcounter{page}{1}

%\copyrightheading{Vol.}{No.}{Year}{Page Nos.}
\copyrightheading{0}{0}{2011}{000--000}

\vspace*{0.88truein}

\alphfootnote

\fpage{1}

\begin{center}\bf
Recovery in quantum error correction for general noise without measurement

\vspace*{0.37truein} \rm \footnotesize
CHI-KWONG LI \\
\vspace*{0.015truein}
{\it Department of Mathematics, College of William \& Mary,
Williamsburg, VA 23187-8795, USA. \\
(Year 2011: Department of Mathematics, Hong Kong University
of Science \& Technology, Hong Kong.)
\rm (Email: ckli@math.wm.edu)}

\vspace*{10pt}
MIKIO NAKAHARA \\
\vspace*{0.015truein}
{\it Research Center for Quantum Computing,
Interdisciplinary Graduate School of Science and Engineering,
and Department of Physics, Kinki University,
3-4-1 Kowakae, Higashi-Osaka, 577-8502, Japan.\\
\rm (Email: nakahara@math.kindai.ac.jp)}

\vspace*{10pt}
YIU-TUNG POON \\
\vspace*{0.015truein}
{\it Department of Mathematics, Iowa State University,
Ames, IA 50011, USA.\\
\rm (Email: ytpoon@iastate.edu)}

\vspace*{10pt}
NUNG-SING SZE\footnote{Corresponding author.}\\
\vspace*{0.015truein}
{\it Department of Applied Mathematics, The Hong Kong Polytechnic University,
Hung Hom, Hong Kong.
\rm (Email: raymond.sze@inet.polyu.edu.hk)}

\vspace*{10pt}
HIROYUKI TOMITA \\
\vspace*{0.015truein}
{\it Research Center for Quantum Computing,
Interdisciplinary Graduate School of Science and Engineering,
Kinki University, 3-4-1 Kowakae, Higashi-Osaka, 577-8502, Japan.\\
\rm (Email: tomita@alice.math.kindai.ac.jp)}

\vspace*{0.225truein}
\publisher{(received date)}{(revised date)}
\end{center}

\vspace*{0.21truein}

\abstracts{
It is known that one can do quantum error
correction without syndrome measurement, which is often done in operator
quantum error correction (OQEC). However, the physical realization could be
challenging, especially when the recovery process involves high-rank
projection operators and a superoperator. We use operator theory
to improve OQEC so that the implementation can
always be done by unitary gates followed by a partial trace operation.
Examples are given to show that our error correction scheme outperforms
the existing ones in various scenarios.
}{}{}

\vspace*{10pt}
\keywords{quantum error correction, operator quantum error correction,
higher rank numerical range, mixed unitary channel}

\vspace*{3pt}
\communicate{to be filled by the Editorial}

\vspace*{1pt}\textlineskip    %) USE THIS MEASUREMENT WHEN THERE IS

\section{Introduction}
\setcounter{equation}{0}

Quantum systems are vulnerable to disturbance from an external environment, which can lead
to decoherence in the system. We have to overcome this difficulty in order
to realize a working quantum
computer and dependable quantum information processing. Quantum error correction (QEC)
\cite{NO,NC,G} is one of the most promising candidates for overcoming decoherence.

QEC proposals to date are separated roughly into two classes: one employs extra ancilla qubits
for error syndrome readout, while the other, called operator quantum error correction (OQEC),
employs high-rank projection operators  based on the Knill-Laflamme result; for example,
see \cite[Theorem 10.1]{NC} and its proof, and also
\cite{kl,KLPL}. There has been strong interest in constructing practical
QEC schemes in actual quantum computing or quantum information
processing.  The major obstacles for the implementation include the following: the syndrome
must be read out by introducing extra ancilla qubits during computing/information processing in the
former case, while realization of high-rank projection operators is
physically challenging in the latter case.

It was shown in \cite{TN} that, for some quantum channels,
there exist different QEC schemes in which no syndrome measurements,
no syndrome readout ancillas and no projection operators were required.
In this scheme,  the recovery   and decoding operations
are combined into a single unitary operation,
and the output state is a direct product of a decoded data qubit state and an encoding
ancilla state. The data qubit state is reproduced without recovering the codeword, and 
moreover, one can see from our result and proofs that
the projection operation in the Knill-Laflamme condition
\cite[Theorem 10.1]{NC} is automatically built into our output state.
The purpose of this paper is to
extend the results in \cite{TN} to general quantum channels. We show that
for any quantum channel
there is a unitary recovery operation for which the output state is a
tensor product of the data
qubit state and an encoding ancilla state.
As a result, a decoding scheme can
be realized by a unitary operation followed by a partial trace operation.
It is worth noting that by a result of Stinespring \cite{dilation}, if the quantum states
are represented by density operators acting on a Hilbert space $\cH$, then
every quantum operation or channel (trace preserving completely positive linear map) can be
realized as a dilation of the density operators to density operators acting on a
Hilbert space $\cK$ followed by a partial trace operation, where $\cK$
is usually of much higher dimension. In our scheme, there is no need to do the
dilation, and only a unitary similarity transform is required. In some examples, one
may  use a permutation similarity transform, or a simple circuit diagram to implement
the unitary similarity transform.
It is also worth noting that there are other automated QEC schemes.
For instant, in the scheme described in \cite{Mermin},
one needs ancillas for error detection, and thus, the number of the
extra qubits is the same as the conventional QECC.
Nevertheless, it still requires additional ancilla qubits whereas our scheme does not.

The rest of the paper is organized as follows. We introduce the basic
notions of QEC and then prove the main theorem in Section 2.
We also give simple examples demonstrating our result
and a simplified proof of a theorem given in \cite{NC}
illustrating that our recovery channel can be used to do correction
for many other channels related to ours.
Section 3 is devoted to summary and discussions.

\section{QEC without Measurement}

Denote by $M_{m,n}$ the set of $m\times n$ complex matrices and
let $M_n := M_{n,n}$ for simplification.
%Denote by $M_n$ the set of $n\times n$ complex matrices.
Let $\Phi:M_n \rightarrow M_n$ be a generalized quantum channel
(i.e., a completely positive linear map without the trace-preserving requirement).
Then a $k$-dimensional subspace $V \subseteq \IC^n$ is a quantum error-correcting code for $\Phi$
if there is a positive scalar $\gamma$
and  a quantum operation $\Psi: M_n \rightarrow M_n$ known as the recovery
channel, such that $\Psi\circ\Phi(\rho) = \gamma \rho$ whenever the state (density matrix)
$\rho$ satisfies $P\rho P = \rho$, where $P$ is the projection operator onto $V$.
A necessary and sufficient condition for the existence of
a quantum  error-correcting code was found by
Knill and Laflamme \cite{kl} (see \cite[Theorem 10.1]{NC}, for example).

\begin{theorem} \label{KL}
Let $\Phi: M_n \rightarrow M_n$ be a quantum channel with the following
operator sum representation
\begin{eqnarray}\label{qc}
\Phi(\rho) = \sum_{j=1}^r F_j \rho F_j^{\dag}\,.
\end{eqnarray}
Suppose $P\in M_n$ is a rank-$k$ orthogonal projection with range space $V$.
The following conditions are equivalent.
\begin{itemize}
\item[{\rm (a)}] $V$ is a quantum error correcting code for $\Phi$.
\item[{\rm (b)}] For $1 \le i, j \le r$, $PF_i^{\dag}F_jP = \lambda_{ij}\, P$
with some complex numbers
$\lambda_{ij}$ so that $\left[ \lambda_{ij} \right]$ is Hermitian.
\end{itemize}
\end{theorem}

In the context of quantum error correction,
the matrices $F_1, \dots, F_r$ in (\ref{qc})
are known as the error operators associated with the channel
$\Phi$; for example see \cite[Chapter 10]{NC}.
The proof of Theorem 10.1 in \cite{NC}
provides a procedure for constructing a
recovery channel $\Psi$ for $\Phi$.
The focus of OQEC schemes will be on constructing and implementing the
recovery channel $\Psi$ for the given channel $\Phi$ without measurement.
However, the recovery channel $\Psi$ may be hard to implement as
the construction involves projection operators and a superoperator.
In this connection,
we show in the following that one can compose the quantum channel $\Phi$
with a unitary similarity transform so that
the output state is a direct sum of the zero operator
and a tensor product of the decoded data qubit state and
an encoding ancilla state. In particular,
a simple construction for recovery operators is proposed
when $n$ is a multiple of $k$, which is often the
case in the context of quantum error correction with
$n$ and $k$ being powers of $2$.

\begin{theorem} \label{main}
Let $\Phi: M_n \rightarrow M_n$ be a quantum channel of the form in (\ref{qc}).
Suppose the equivalent conditions in Theorem \ref{KL} hold and
$P=WW^{\dag}$ with $W^{\dag}W = I_k$ so that a density matrix $\rho \in M_n$ satisfying
$P\rho P = \rho$ has the form $W\tilde \rho W^{\dag}$ with $\tilde\rho \in M_k$.
Then there is an $R \in U(n)$ and a positive definite matrix $\xi \in M_q$
with $q \le \min\{r,n/k\}$ such that
for any density matrix $\tilde \rho \in M_k$ and $\rho =
W\tilde \rho W^{\dag}\in M_n$, we have
$$R^{\dag}\,\Phi(\rho)\,R = (\xi \otimes \tilde \rho) \oplus 0_{n-qk}.$$
In particular, if $k$ divides $n$ so that $M_n$ can be regarded as
$M_{n/k}\otimes M_k$, then
$$R^{\dag}\,\Phi(\rho)\,R = \tilde \xi \otimes \tilde \rho \quad \hbox{ with } \quad
\tilde \xi = \xi \oplus 0_{n/k-q}$$
and a recovery channel can be constructed as the map
$\Psi: M_n \rightarrow M_n$ defined by
$$\Psi(\rho') = W(\, \tr_1(R^\dag \rho' R)\, )W^{\dag},$$
where
$\tr_1$ stands for the partial trace over the encoding ancilla Hilbert space.
If $\Phi$ is trace preserving, i.e.,
$\sum_{j=1}^r F_j^{\dag}F_j = I_n$, then $\tr\, \xi = 1$ so that
$\Psi$ is also trace preserving.
\end{theorem}

\noindent\bf Proof. \rm
Suppose the equivalent conditions in Theorem \ref{KL} hold,
i.e.,
$PF_i^{\dag}F_jP = \lambda_{ij} P$ for some $\lambda_{ij} \in \IC$.
Notice that $\Lambda = [\lambda_{ij}]$ is an $r\times r$ positive semi-definite matrix.
Suppose $\Lambda$ has rank $q$.
Then there is a $U = [u_{ij}] \in U(r)$ and a
positive semi-definite matrix
$\hat \xi = \left[\hat \xi_{ij}\right]\in M_r$ such that $U^\dag \Lambda U = \hat \xi$
and $\hat \xi_{ij} = 0$ for all $q < i \le r$ or $q \le j \le r$. Equivalently,
$\hat \xi = \xi \oplus 0_{r-q}$
for some positive definite matrix $\xi = \left[\xi_{ij}\right] \in M_q$. Define
$$\tilde F_j = \sum_{i=1}^r u_{ij} F_i \quad \hbox{ for }  j = 1,\dots, r.$$
Let
$$F = \left[\matrix{ F_1  & F_2
& \cdots & F_r } \right]$$ be an $n\times rn$ matrix obtained by
a juxtaposition of $\{F_j\}_{1 \le j \le r}$
in the given order. Similarly, write
$\tilde F = \left[\matrix{ \tilde F_1  & \tilde F_2
& \cdots & \tilde F_r } \right].$
Then
$\tilde F = F( U \otimes I_n)$
and for any $\rho \in M_n$,
\begin{eqnarray*}
\Phi(\rho)
&=& \sum_{j=1}^r F_j \rho F_j^{\dag}
= F (I_r \otimes \rho) F^{\dag}
= F (U \otimes I_n) (I_r \otimes \rho)(U \otimes I_n)^{\dag} F^{\dag} \\[1mm]
&=& \tilde F (I_r \otimes \rho) \tilde F^{\dag}
= \sum_{j=1}^r \tilde F_j \rho \tilde F_j^{\dag}.
\end{eqnarray*}
So $\Phi(\rho) = \sum \tilde F_j \rho \tilde F_j^{\dag}$
is another operator sum representation of $\Phi$.
Furthermore,
$$P\tilde F_i^\dag \tilde F_j P
= \sum_{k,l=1}^r u^*_{ki} u_{lj} PF_{k}^\dag F_{l} P
= \sum_{k,l=1}^r u^*_{ki} u_{lj} \lambda_{kl} P
= \hat \xi_{ij} P \quad\hbox{for all}\quad i,j = 1,\dots,r.
$$
Without loss of generality, we may assume that $\tilde F_j = F_j$ and
$PF_i^\dag F_j P =  \xi_{ij} P$ for all $1\le i,j\le q$.
Furthermore, replace the matrix $F$ defined above by
$F = \left[\matrix{ F_1  & F_2
& \cdots & F_q } \right].$
Since $P = WW^{\dag}$ with $W^{\dag}W = I_k$, it follows that
$$ W^{\dag} F_i^{\dag}F_j W = \xi_{ij} I_k \quad \hbox{ which is equivalent to } \quad
(I_q \otimes W)^{\dag} F^{\dag}F (I_q \otimes W) = \xi \otimes I_k. $$
Define
an $n \times qk$ matrix
$$R_1 %= F (D^{-1/2} \otimes W)
= F(I_q \otimes W)(\xi^{-1/2} \otimes I_k).$$
Then $R_1^{\dag}R_1 = I_{qk}$.
Take an $n \times (n-qk)$ matrix $R_2$
such that $R = \left[\matrix{ R_1 & R_2 }\right] \in U(n)$. Then
$$R^{\dag} F (I_q \otimes W)
= R^{\dag} R_1 (\xi^{1/2} \otimes I_k)
= \left[\matrix{
\xi^{1/2} \otimes I_k \cr
0 } \right].$$
Now for any $\rho\in M_n$ with $P\rho P = \rho$, there exists $\tilde \rho
\in M_k$ such that
$\rho = W \tilde \rho W^{\dag}$.
Since $W^{\dag}F_j^\dag  F_jW = \hat \xi_{jj}I_k= 0$ and hence $F_jW = 0$
for all $j > q$, $\Phi(\rho)$ can be written as
$$\Phi(\rho)
= \sum_{j = 1}^r F_j (W\tilde \rho W^{\dag}) F_j^{\dag}
= \sum_{j = 1}^q F_j (W\tilde  \rho W^{\dag}) F_j^{\dag}
= F (I_q \otimes (W\tilde  \rho W^{\dag})) F^{\dag}.$$
%= F (I_q \otimes W)(I_q \otimes \rho)(I_q\otimes W^{\dag}) F^{\dag}.$$
It follows that
\begin{eqnarray*}
R^{\dag} \Phi(\rho) R
&=&R^{\dag}F (I_q \otimes W)(I_q \otimes \tilde \rho)(I_q\otimes W^{\dag}) F^{\dag}R \\[1mm]
&=& \left[\matrix{
\xi^{1/2} \otimes I_k \cr
0 } \right]
(I_q \otimes \tilde \rho)
\left[\matrix{
\xi^{1/2} \otimes I_k &
0 } \right]
= \left[ \matrix{
\xi \otimes \tilde \rho & 0 \cr 0 & 0 }\right].
\end{eqnarray*}
Now if $k$ divides $n$, we have shown that
$
\Psi \circ \Phi(\rho) = W \left[\tr_1( R^{\dagger}\Phi(\rho) R)\right]
W^{\dagger} = W\tilde \rho W^{\dagger} =\rho$ as promised.

Finally, to see that $\sum_{j=1}^q \xi_{jj} = 1$ if $\sum_{j=1}^q F_j^{\dag}F_j = I_n$, note that
$$P = P\left(\sum_{j=1}^r F_j^{\dag}F_j\right)P = \sum_{j=1}^r PF_j^{\dag}F_jP
= \left(\sum_{j=1}^r \hat \xi_{jj} \right)P = \left(\sum_{j=1}^q \xi_{jj} \right)P.$$ The result follows.
\qed

\medskip
Note that we have shown that if a channel $\Phi$ is correctable, its action
on the states $\rho$ satisfying $P\rho P = \rho$ is very simple, namely,
$$
\Phi(\rho) = R[(\xi\otimes (W^{\dag}\rho W)) \oplus 0]R^{\dag}.
$$
As a result, we can easily recover $\rho$
from $\Phi(\rho)$.
It is worth pointing out several features of our scheme.

First, it is known that a recovery channel is a (trace preserving) completely
positive linear map, and such a map can always be realized by a dilation of the
basic system to a much larger system, followed by a compression \cite{dilation}. In contrast,
our scheme does not require a dilation of the basic system to a larger system.
%In the previously described scenario, we need only work on the quantum computing
%device which can handle $n$ qubits without requiring any modification of the device
%so that it can handle more qubits.

Second, suppose one considers the algebra generated by the error operators
of the quantum channel describing the decoherence that may affect the
quantum computing device, and one obtains a decomposition
of the algebra as $(M_s \otimes I_r) \oplus \cA$.
Then one has a noiseless subsystem of dimension $r$ so that
a state of the form $(\xi \otimes \rho) \oplus 0$  will be mapped
to a state of the form $(\tilde \xi \otimes \rho) \oplus 0$ in which
the data state $\rho \in M_r$ is not affected by the quantum channel at all;
see \cite{KLPL,Kempe}.
Our result shows that as long as a QECC of dimension $r$ exists,
one can construct a unitary operation $R$
such that when one encodes a data state $\rho \in M_r$ by
$W \rho W^\dag$, where $WW^\dag$ is the orthogonal projection with
QECC as its range space,
then the quantum channel will send the encoded state to
$R(\xi \otimes \rho)R^{\dag}$.
Thus, one can recover $\rho$ by
  a unitary operation and discarding of a subsystem.
%, and then apply the inverse of the unitary similar transform.
Hence, our encoding and decoding scheme strongly resembles  the
noiseless subsystem approach, but the use of the algebra generated by
the error operators is unnecessary. In fact, if we consider the
mixed unitary channel $\rho \mapsto (\rho + U\rho U^{\dag})/2$ with
diagonal unitary $U = \diag\left(1,-1,i,-i\right)$,
then the algebra generated by the error operators
$I/\sqrt{2}$ and $U/\sqrt{2}$
is the algebra of diagonal matrices. Thus, there is no non-trivial noiseless subsystem.
Nonetheless we can find a 2-dimensional QECC (one data qubit), and apply our scheme as shown
in the following example.

{\bf Example 1}
Consider a mixed unitary channel $\Phi(\rho) =  (\rho + U\rho U^{\dag})/2$ with
diagonal unitary $U = \diag\left(1,-1,i,-i\right)$.
One can find a 2-dimensional QECC,
which is spanned by the codewords
$$|\bar 0\ra = (|00\ra+|01\ra)/\sqrt{2}
\quad\hbox{and}\quad
|\bar 1\ra = (|10\ra + |11\ra)/ \sqrt{2}.$$
In this case, the corresponding projection operator is given by $P = WW^\dag$ with
$$W = \frac{1}{\sqrt 2} \left[\matrix{ 1 & 1 & 0 & 0 \cr 0 & 0 & 1  & 1 }\right]^\dag.$$
Following the proof of Theorem \ref{main}, one can construct
the recovery operator $R$ as
$$R = \frac{1}{\sqrt 2} \left[\matrix{
1 & 0 & 1  & 0  \cr
1 & 0 & -1 & 0 \cr
0 & 1 & 0  & i \cr
0 & 1 & 0  & -i }\right].$$
Then one can check that for a codeword $\rho = W\tilde \rho W^\dag$ with $\tilde \rho \in M_2$, we have
$$R^\dag \Phi(\rho) R = \frac{1}{2}\, I_2 \otimes \tilde \rho.$$

Third, even though we cannot say that our scheme
is always better than other QEC schemes,
there are examples of noisy channels in which our scheme is
simple to implement; see our recent works \cite{QECC2,QECC3}.
Furthermore, comparing
our scheme with the one in the proof of the Knill-Laflamme theorem,
one can certainly see the advantage in our result as illustrated in the
examples below.
%{\bf In the subsequent discussion,  we have to show the scheme of
%Knill-Laflamme, and compare with ours in our examples.}

Finally, in Theorem 3 we illustrate that one can use the same encoding and decoding
scheme to deal with new quantum channels obtained from the given one
whenever the error operators are obtained from linear combinations of the old
ones. This allows us to deal with quantum channels with error operators
chosen from an infinite set. %; see our recent work.
Theorem \ref{main} was demonstrated for three-, five- and nine-qubit quantum
error correcting codes explicitly in \cite{TN}, see also \cite{B,7qubit}.
It is instructive to work out the simplest example of the three-qubit bit-flip QEC to clarify
the theorem in the following.

\bf Example 2 \rm
We take a pure state data qubit to simplify the notation.
A one-qubit data state
$|\psi_0\ra = \alpha|0\ra + \beta |1\ra$ is encoded with two encoding
ancilla qubits as
$|\psi\ra = \alpha |000\ra	 + \beta|111\ra$,
which is an element of the code space $V$.
The projection operator is
$$P = \diag(1, 0, 0, 0, 0, 0, 0, 1),$$
which is also written as $P = WW^{\dag}$, where
$$W =\left[\matrix{
1 & 0 & 0 & 0 & 0 & 0 & 0 & 0 \cr
0 & 0 & 0 & 0 & 0 & 0 & 0 & 1 }\right]^\dag.$$
Evidently,  $W^{\dag}W = I_2$. Let
\begin{eqnarray}\label{rho0}
\tilde{\rho} = |\psi_0\ra \la \psi_0| 	=
\left[\matrix{|\alpha|^2 & \alpha \beta^* \cr \alpha^* \beta & |\beta|^2\cr
}\right].
\end{eqnarray}
The encoded state is then
\begin{eqnarray}\label{rho}
\rho = W\tilde{\rho} W^{\dag} =
\left[\matrix{
|\alpha|^2 & 0 & 0 & 0 & 0 & 0 & 0 & \alpha \beta^* \cr
0 & 0 & 0 & 0 & 0 & 0    & 0 & 0 \cr
  & & \vdots & & & \vdots & & \cr
0 & 0 & 0 & 0 & 0 & 0    & 0 & 0 \cr
\alpha^*\beta & 0 & 0 & 0 & 0 & 0 & 0 & |\beta|^2 \cr}\right].
\end{eqnarray}
The bit-flip quantum channel is defined as
$$\Phi(\rho) = \sum_{i=0}^3 F_i \rho F_i^{\dag},$$
where
\begin{equation}\label{F}
%\qquad
F_0 =\sqrt{p_0}\, I_2 \otimes I_2 \otimes I_2,
\quad
F_1 =\sqrt{p_1}\, \sigma_x \otimes I_2 \otimes I_2,
\quad
F_2 =\sqrt{p_2}\, I_2 \otimes \sigma_x \otimes I_2,
\quad
F_3 =\sqrt{p_3}\, I_2 \otimes I_2 \otimes \sigma_x,
\end{equation}
with $p_0+\cdots + p_3 \le 1$. Here $\sigma_i$ is the $i$-th Pauli matrix.
It is easy to verify that
$W^{\dag}F_i^{\dag}F_j W = p_i \delta_{ij} I_2$.
% is equal to $p_i I_2$ if $i = j$ and equal to zero if $i\ne j$.
Following the proof of Knill-Laflamme's result,
see \cite[Theorem 10.1]{NC} for example,
one can construct the recovery channel
$\Psi: M_8 \to M_8$ given by
$$\Psi(\rho') = P \rho' P + (I_8 - P) \rho' (I_8 - P).$$
Then we have $\Psi \circ \Phi(\rho) = \rho$
for all codewords $\rho = W\tilde \rho W^\dag$.
However, the quantum channel $\Psi$ is hard to
implement as it involves projection operators $P$ and $(I_8 - P)$
and, moreover, $\Psi$ is a superoperator.
On the other hand, notice that
$$(I_4 \otimes W)^{\dag}F^{\dag}F(I_4\otimes W) = \xi\otimes I_4,$$
where
$F = [F_0\ F_1\ F_2\ F_3]$
and $\xi = \diag(p_{0}, p_{1}, p_{2}, p_{3})$.
Following the proof of Theorem \ref{main},
let $R_1 = F(I_4\otimes W)(\xi^{-1/2} \otimes I_2)$.
Direct computations yield
\begin{equation}\label{R}
R_1 = E_{11} + E_{27}
+ E_{35} + E_{44}
+ E_{53} + E_{66}
+ E_{78} + E_{82},
\end{equation}
where $\{E_{11}, E_{12}, \dots, E_{88}\}$ is the standard basis for $M_8$.
Then $R_1^{\dag}R_1 = I_8$.
The matrix $R_2$ in the proof of Theorem \ref{main}
is vacuous since $R_1$ is unitary by itself. We
denote $R_1$ as $R$ hereafter. Note that
$$R^{\dag}
F(I_4 \otimes W) = \xi^{1/2}\otimes I_2.$$
For a codeword $\rho = W\tilde{\rho} W^{\dag}$, we have
\begin{equation}\label{phirho1}\Phi(\rho) =
\sum_{j=0}^3 F_j(W\tilde{\rho} W^{\dag})F_j^{\dag} =
F(I_4 \otimes (W\tilde{\rho} W^{\dag}))F^{\dag}.\end{equation}
It follows that
\begin{equation}\label{phirho2}R^{\dag} \Phi(\rho) R =
R^{\dag}F (I_4\otimes W)(I_4\otimes \tilde{\rho})(I_4 \otimes W^{\dag})F^{\dag}R = \xi \otimes \tilde{\rho}.\end{equation}
Now the decoded data state $\tilde{\rho}$  appears in the output with no syndrome measurements nor explicit projection.
It %has to point
should be pointed out that
the unitary operation $R$ in (\ref{R}) is independent of the choice of
nonnegative numbers $p_j$.
A simple encoding and recovery circuit for 3-qubit bit-flip channel,
which encodes and recovers an arbitrary $1$ qubit state
with two ancilla qubits, was presented in \cite{TN}.
We also note  {\it en passant} that this QEC was
obtained in \cite{B} from different viewpoint based on classical
error correction.

Recently, using the same scheme and the techniques
of higher rank numerical range, we have shown in \cite{QECC2}
that there is a quantum error correction
which suppresses fully correlated errors of the form
$\sigma_i^{\otimes n}$.
It has been proved that $n$ qubit codeword encodes
(i) $(n-1)$ data qubit states when $n$ is odd
and (ii) $(n-2)$ data qubit states when $n$ is even.
Furthermore, it has been proved that
one cannot encode $(n-1)$ qubits for even $n$.
This shows that our QEC is optimal in this setting.

In \cite[Theorem 10.2]{NC}, the authors showed that the recovery operation
constructed for a given quantum channel $\Phi$
in their Theorem 10.1 can be used to correct error of other
channels whose error operators are linear combinations of those of $\Phi$.
In the following, we show that the recovery channel constructed
in Theorem \ref{main}
above has the same property. In particular, if $R$ is
the unitary matrix constructed for $\Phi$ in Theorem \ref{main},
then $R^{\dag} \tilde \Phi(\rho) R$ always have the desired direct sum
structure.

\begin{theorem}\label{2.3}
Suppose $R$ is the unitary matrix given in Theorem \ref{main}.
If $\tilde \Phi$ is another quantum channel $\tilde \Phi(\rho) = \sum \tilde F_j \rho \tilde F_j^\dag$,
where the error operators $\tilde F_j$'s are linear combinations of $F_j$'s,
then there is a positive definite $\tilde \xi$ such that
for any density matrix $\tilde \rho \in M_k$
and $\rho = W\tilde \rho W^{\dag}\in M_n$, we have
$$R^{\dag}\,\tilde \Phi(\rho)\,R = (\tilde \xi\otimes \tilde \rho) \oplus 0.$$
\end{theorem}

\noindent\bf Proof. \rm
We use the same notations as in Theorem \ref{main}.
Suppose $\tilde F_j$'s are linear combinations
of $F_i$'s, i.e.,
$$\tilde F_j = \sum_{i = 1}^r t_{ij} F_i \quad \hbox{
for } \quad j = 1,\dots, s.$$
Recall that $F_j W = 0$ for all $j > q$.
Then
$\tilde F_j W = \sum_{i = 1}^q t_{ij} F_i W$ for all $j =1 ,\dots, s$.
Define a $q \times q$ matrix $T = \left[ t_{ij} \right]_{1\le i,j\le q}$. 
For any codeword $\rho = W\tilde \rho W^\dag$,
by a similar argument as in the proof of Theorem \ref{main}, 
one can see that
$$\tilde \Phi(\rho)
= F (TT^\dag \otimes W\tilde \rho W^\dag) F^\dag.
$$
Then
$$R^\dag \tilde \Phi(\rho) R
= (\tilde \xi\otimes \tilde \rho) \oplus 0
\quad\hbox{where}\quad
\tilde \xi = \xi^{1/2} TT^\dag \xi^{1/2}.$$
\vskip-.20in \hspace{115mm}
\qed

\medskip
Note that by Theorem 3, for a given quantum channel
$\Phi$ in operator sum form with error operators
$\{F_1, \ldots, F_r\}$, one may choose a set of operators
$\{E_1, \ldots , E_m\}$, where $m \leq r$, in the simplest from so that
the set has the same linear span as $\{F_1, \ldots, F_r\}$.
Then construct the new channel
$\tilde \Phi(\rho) = E_1\rho E_1^{\dagger} + \ldots + E_m \rho E_m^{\dagger}$
and the recovery operation. The resulting recovery channel for
$\tilde{\Phi}$ corrects the original channel $\Phi$.

\bf Example 3 \rm
To illustrate the above remark,
%consider the three-qubit bit-flip channel after Theorem 2.
%In the construction of the recovery channel $\Psi$, one may assume that all
%the values $p_0, p_1, p_2, p_3$ in (4) equal $p$ to simplify the
%construction.
consider
a quantum channel $\Phi: M_8 \to M_8$ defined as
$$\Phi(\rho) = \sum_{i=0}^3 \tilde F_i \rho \tilde F_i^{\dag},$$
where
$$\begin{array}{ll}
\tilde F_0 =\sqrt{\tilde p_0} I_2 \otimes I_2 \otimes I_2, &
\tilde F_1 = \sqrt{\tilde p_1} e^{it_1\sigma_x} \otimes I_2 \otimes I_2, \\[3mm]
\tilde F_2 =\sqrt{\tilde p_2} I_2 \otimes e^{it_2\sigma_x} \otimes I_2, \qquad &
\tilde F_3 =\sqrt{\tilde p_3} I_2 \otimes I_2 \otimes e^{it_3\sigma_x},
\end{array}$$
with $t_1,t_2,t_3\in \IR$ and probability $\tilde p_j$ such that $\sum_{j=0}^3 \tilde p_j \le 1$.
Since $e^{it \sigma_x} = \cos t\ I + i\sin t\ \sigma_x$, we see that $\tilde F_0, \dots, \tilde F_3$
are linear combinations of $F_0, \dots, F_3$ of the three-qubit channel
introduced in Example 2.
Thus, the recovery channel $\Psi$
%(or the simplified version using $p_0 = p_1 = p_2 = p_3 = p$)
constructed previously can be used for this channel also.
%%
%
%\medskip
%Of course, one can also give the details of the connection between
This channel and the three-qubit channel introduced previously
%after Theorem \ref{main}
are related as follows.
%Notice that if we assume that $p_i = p$ $(0 \leq i \leq 3)$, then
%\iffalse
The error operators $\tilde F_j$ of the present channel
are linear combinations of $F_j$, defined in (\ref{F}).
More precisely,
%\fi
$$\tilde F_j = \sum_{i = 0}^3 t_{ij} F_i \quad\hbox{for}\quad j = 0,1,2,3,$$
where
$$T = \left[ t_{ij} \right]
= %\frac{1}{\sqrt{p}}
\left[\matrix{
\sqrt{  {\tilde p_0}/{p_0}}
& \sqrt{ {\tilde p_1}/{p_0}}\, \cos t_1
& \sqrt{ {\tilde p_2}/{p_0}}\, \cos t_2
& \sqrt{ {\tilde p_3}/{p_0}}\, \cos t_3 \cr
0 &
i \sqrt{ {\tilde p_1}/{p_1}}\, \sin t_1 & 0 & 0 \cr
0 & 0 &
i \sqrt{ {\tilde p_2}/{p_2}}\, \sin t_2 & 0 \cr
0 & 0 & 0 &
i \sqrt{ {\tilde p_3}/{p_3}}\, \sin t_3
}\right].$$
%where we assumed that
%\iffalse
%$F_i$ are of the form of the first example with
%\fi
%$p_i = p$ $(0 \leq i \leq 3)$ to simplify the matrix $T$.
\iffalse
By Theorem \ref{2.3}, the unitary operation $R$ defined in (\ref{R})
together with the partial trace operation
also gives rise to a recovery channel for the channel $\tilde \Phi$.
%To see this, suppose a one-qubit state
%$|\psi_0\ra = \alpha|0\ra + \beta |1\ra$ is encoded with two ancilla %qubits as
%$|\psi\ra = \alpha |000\ra	 + \beta|111\ra$,
%which is an element of the code space $V$ .
\fi
Define $\tilde{\rho}$ and $\rho$ as in (\ref{rho0}) and (\ref{rho}).
Then, similar to the computation in (\ref{phirho1}) and (\ref{phirho2}), we have 
$$R^\dag\, \Phi(\rho)\, R = \tilde \xi \otimes \tilde \rho,$$
where $\tilde \xi$ is defined by $\xi^{1/2} TT^\dag \xi^{1/2}$,
which is equal to
$$\hspace{-3mm}
\left[ \matrix{
\tilde p_0 + \tilde p_1 \cos^2 t_1 + \tilde p_2 \cos^2 t_2 + \tilde p_3 \cos^2 t_3 &
- i  \tilde p_1 \cos t_1 \sin t_1 &
- i  \tilde p_2 \cos t_2 \sin t_2 &
- i  \tilde p_3 \cos t_3 \sin t_3 \cr
i  \tilde p_1 \cos t_1 \sin t_1 &
\tilde p_1 \sin^2 t_1 &
0 & 0 \cr
i  \tilde p_2 \cos t_2 \sin t_2 &
0 & \tilde p_2 \sin^2 t_2 & 0 \cr
i  \tilde p_3 \cos t_3 \sin t_3
& 0 & 0 & \tilde p_3 \sin^2 t_3
}\right].
$$

\section{Summary}

We have shown that QEC without syndrome measurements is possible, such that the output state
is a tensor product of a decoded data qubit state and an encoding ancilla state. The recovery operation is combined with
the decoding operation, so that both are implemented by a unitary operation. We gave a constructive
proof that there always exists such a unitary operator for a given quantum channel. We also
prove a result analogous to  \cite[Theorem 10.2]{NC},
namely, we show that the recovery operation
constructed for a quantum channel $\Phi$ in our main theorem is automatically a recovery
channel for a channel whose error operators are linear combinations of those of $\Phi$.

Most of the QECs proposed so far are based on the code space. A data
qubit state $\tilde{\rho}$ is encoded as
$\rho$ and then a noisy quantum channel $\Phi$ is applied on $\rho$.
The encoded state is recovered first and subsequently
the decoding operation is applied to extract the qubit state $\tilde{\rho}$.
Note, however, that what we need eventually is $\tilde{\rho}$ and not $\rho$.
Our QEC is arranged in such a way that the output is a tensor product
of $\tilde{\rho}$ and an encoding ancilla state so that $\tilde{\rho}$
is obtained %simply tracing out the ancilla Hilbert space
without any syndrome measurement or projection. It follows from our result
that QEC can be accomplished by applying a unitary gate followed by a
partial trace operation.
%Such a scheme requires fewer
%ancilla qubits and provides a more efficient %encoding and
%recovery operations.

\section*{Acknowledgments}
This research began at the 2nd Workshop on Quantum Information Science,
supported by a HK RGC grant
and hosted by the Hong Kong Polytechnic University in January 2011.
The research of Li was supported by a USA NSF grant, a HK RGC grant,
the 2011 Fulbright Fellowship, and the 2011 Shanxi 100 Talent Program.
He is an honorary professor of University of Hong Kong,
Taiyuan University of Technology, and Shanghai University.
The research of Nakahara and Tomita was supported
by ``Open Research Center'' Project
for Private Universities: matching fund subsidy from MEXT
(Ministry of Education, Culture, Sports, Science and Technology).
The research of Poon was supported by a USA NSF grant and a HK RGC grant.
The research of Sze was supported by HK RGC grants.
Finally, the authors wish to thank the two referees for their valuable comments and suggestions 
which led to a significant improvement of the manuscript.


\section*{References}
\begin{thebibliography}{WWW}

\bibitem{NO} M.~Nakahara and T.~Ohmi, 
Quantum Computing. From Linear Algebra to Physical Realizations,
CRC Press, New York, 2008.

\bibitem{NC}
M.~A.~Nielsen and I.~L.~Chuang,
Quantum Computation and Quantum Information, 
Cambridge University Press, 2000.

\bibitem{G}
F.~Gaitan, 
Quantum Error Correction and Fault Tolerant Quantum Computing, CRC Press,
New York, 2008.

\bibitem{kl} E. Knill and R. Laflamme,
Phys. Rev. A {\bf 55}, 900 (1997).

\bibitem{KLPL}
D.W. Kribs, R. Laflamme, D. Poulin, M. Lesosky,
Quant. Inf. \& Comp., {\bf 6}, 383-399 (2006).


\bibitem{TN} H.~Tomita and M.~Nakahara, 
eprint arXiv:quant-ph/1101.0413.


\bibitem{dilation} W.F. Stinespring, 
%Positive functions on $C^*$-algebras, 
Pro. Amer. Math. Soc. {\bf 6}, 211--216 (1955).


\bibitem{Mermin}
N.D. Mermin, Quantum computer science: an introduction,
Cambridge University Press, 2007.

\bibitem{Kempe}
J. Kempe,
S\'{e}minaire Poincar\'{e} {\bf 2}, 1-29 (2005).


\bibitem{QECC2}
C.-K. Li, M. Nakahara, Y.-T. Poon, N.-S. Sze, and H. Tomita,
Phys. Lett. A {\bf 375}, 3255-3258 (2011).

\bibitem{QECC3}
C.-K. Li, M. Nakahara, Y.-T. Poon, N.-S. Sze, and H. Tomita,
Phys. Rev. A {\bf 84}, 044301 (2011).

\bibitem{B} S.~L.~Braunstein, 
eprint arXiv:quant-ph/9603024.

\bibitem{7qubit}  
R.~Laflamme, C.~Miquel, J.~P.~Paz and W.~H.~Zurek,
Phys. Rev. Lett. {\bf 77}, 198 (1996).



\end{thebibliography}
\end{document}